\documentclass{PoS}
\usepackage{mkPhys}

\title{Charmonium Spectroscopy (X,Y,Z) at the B Factories }

\ShortTitle{Charmonium Spectroscopy (X,Y,Z) at the B Factories }

\author{\speaker{Michal Kreps}%
         \thanks{On behalf of the Belle and \babar collaborations.}
        \\Karlsruhe Institute for Technology\\
        E-mail: \email{michal.kreps@kit.edu}}

\abstract{
Since 2003 several states in the charmonium mass region were
discovered. While in the conventional $c\overline{c}$ spectrum
some states are missing, the number of states observed up to
now is larger than empty spaces in the $c\overline{c}$ spectrum.
This, together with other difficulties to explain observed
states as a $c\overline{c}$ mesons triggered discussions on
a possible exotic interpretations. In this proceedings we present current
experimental status from B-factories of the so called $X$, $Y$ and $Z$ states.
}

\FullConference{12th International Conference on B-Physics at Hadron Machines - BEAUTY 2009\\
		 September 07 - 11 2009\\
		 Heidelberg, Germany}

\begin{document}

\section{Introduction}
The charmonium ($c\overline{c}$) spectrum is thought to be
well understood with a good agreement between theory and
experiment. With current generation of B-factories also
search for an unobserved but predicted states began. First
state was observed in 2003 by Belle experiment
\cite{Choi:2003ue} and is known
now as $X(3872)$. In a following years, many other states were
observed. Very quickly it become clear that those states are
difficult to explain as conventional $c\overline{c}$ mesons
given our understanding of the $c\overline{c}$ spectra. All
those mesons have in  common a
mass in charmonium region and a rather narrow width. While
being above the threshold for a decay
to a pair of open charm mesons, they were observed in the decays
containing \Jpsi in a final state. The states are named $X$, $Y$
and $Z$ which reflects the unknown nature of them. 

The difficulty in fitting observed status to the charmonium
spectrum in connection with few other properties discussed
later triggered
speculations that at least some of the observed states might
be candidates for exotic mesons. Those include variety of
options like molecule of two loosely bound mesons,
different variations of four quark states, hybrids or
glueballs. 
In this write-up we present the current status and developments
in experimental studies of the $X$, $Y$ and $Z$ mesons at
B-factories experiments Belle and \babar.

\section{The $X(3872)$}

The $X(3872)$ is best known and studied out of the puzzling
$XYZ$ states. It was observed in 2003 in the $B^+$ decays to
$X(3872)K^+$ with $X(3872)$ decaying to $\Jpsi\pi^+\pi^-$
\cite{Choi:2003ue}. The state was subsequently confirmed in B
decays by \babar experiment \cite{Aubert:2004fc} and in
$p\overline{p}$ production
by Tevatron experiments CDF \cite{Acosta:2003zx} and D\O{}
\cite{Abazov:2004kp}. Already first measurements yielded 
mass in the proximity of $D^0\overline{D}^{*0}$ threshold. From
the subsequent observation of decays $X(3872)\rightarrow
\Jpsi\gamma$ \cite{jpsigamma} and the angular analysis performed by CDF
experiment \cite{Abulencia:2006ma}, the possible quantum numbers were reduced to two
options of $J^{PC}=1^{++}$ or $2^{-+}$. In
addition,  decays to $D^0\overline{D}^{*0}$ were observed
\cite{Gokhroo:2006bt,Aubert:2007rva},
but the mass measured in this decay mode was about $3$ MeV above
the mass measured in the $\Jpsi\pi^+\pi^-$ decay.

\begin{figure}
\centering
\includegraphics[height=0.23\textwidth]{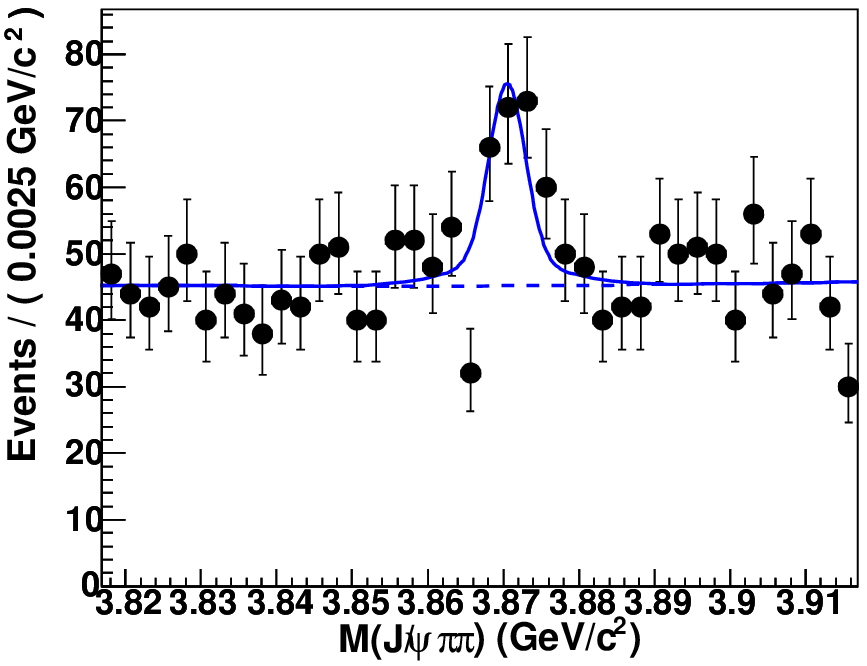}
\hspace{0.05\textwidth}
\includegraphics[height=0.26\textwidth]{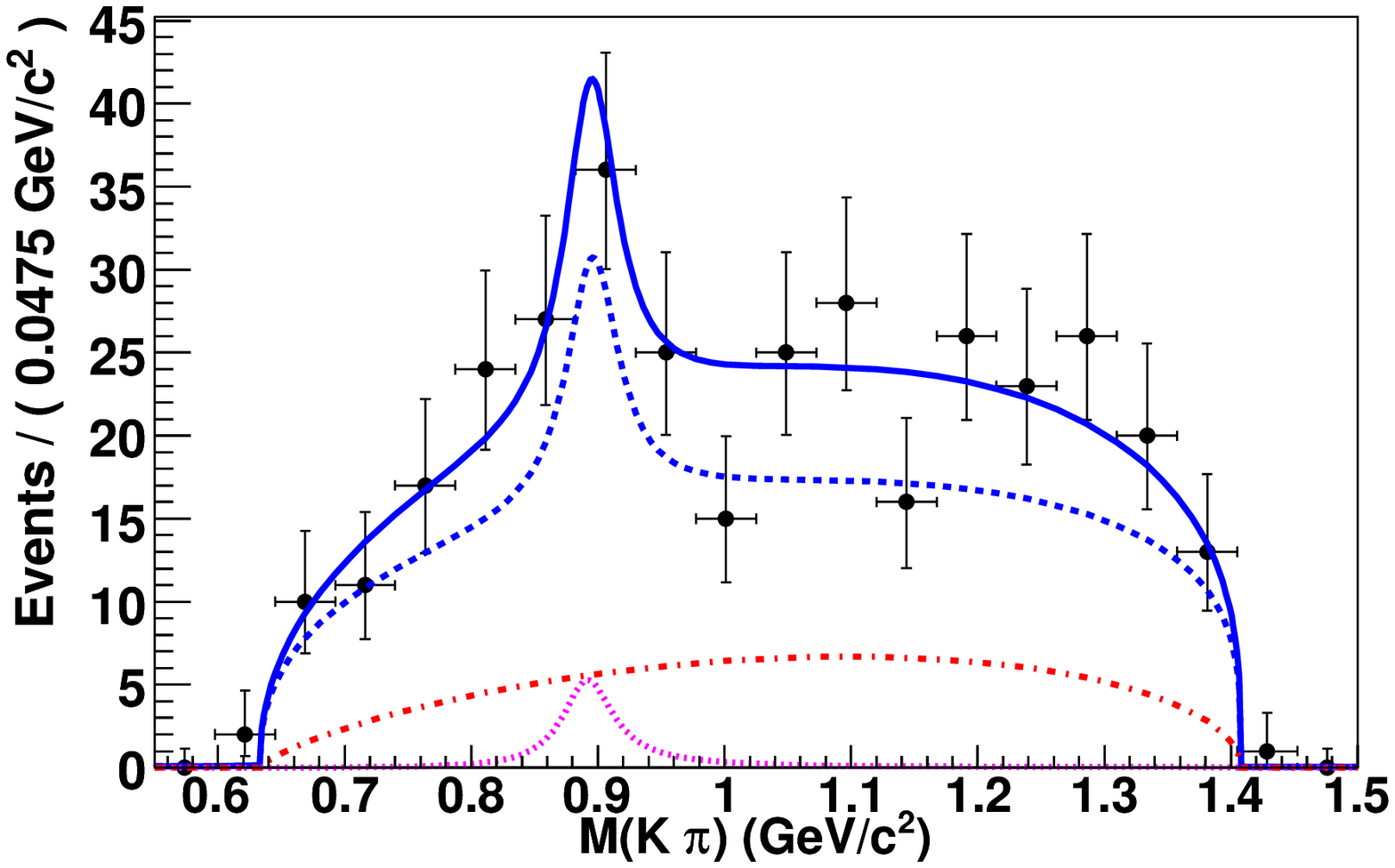}
\caption{The $\Jpsi\pi^+\pi^-$ invariant mass distribution
(left) and $K^+\pi^-$ invariant mass distribution (right)
for $B^0\rightarrow \Jpsi\pi^+\pi^- K^+\pi^-$ from Belle. In
the $K^+\pi^-$ distribution only events in $X(3872)$ region are used. The full line 
shows the fit projection, the dashed blue line
represents background, the dashed-dotted shows non-resonant
decays with $X(3872)$ signal and the dotted curve shows
resonance decays with $X(3872)$ signal. }
\label{fig:belle_x3872_kpi}
\end{figure}

Several new measurements were performed in the past year by
both B-factory experiments. Using 657 million
$B\overline{B}$ pairs, Belle experiment updated
the mass measurement in the $\Jpsi\pi^+\pi^-$ final states
\cite{:2008te}. Combining $B^+\rightarrow X(3872) K^+$ and
$B^0\rightarrow X(3872) K_s$ samples together they measure
$M(X(3872))=3871.46 \pm 0.37 \pm 0.07$ \mevcc. The
difference of the masses measured separately in $B^+$ and
$B^0$ decays is $\delta M=0.18 \pm 0.89 \pm 0.26$ \mevcc.
The analogous result from \babar experiment based on the 455
million $B\overline{B}$ pairs gives $M(X(3872))=3871.4 \pm 0.6 \pm
0.1$ \mevcc and $\delta M=2.7 \pm 1.6 \pm 0.4$ \mevcc
\cite{Aubert:2008gu}. In
both cases, the measured mass of the $X(3872)$ is slightly below
$D^0\overline{D}^{*0}$ threshold, but mass above
the $D^0\overline{D}^{*0}$ threshold cannot be excluded.
The updated study of Belle collaboration adds in addition
first observation of the $X(3872)$ in the $B^0\rightarrow
X(3872) K^+\pi^-$ decay. In Fig.~\ref{fig:belle_x3872_kpi}
we show $\Jpsi\pi^+\pi^-$ and $K^+\pi^-$ invariant mass
distributions. The fit of the $K^+\pi^-$ mass distribution
including possibility of resonant decays $B^0\rightarrow X(3872)
K^{*0}$ as well as non-resonant $B^0\rightarrow X(3872)
K^+\pi^-$ reveals that most of the $X(3872)$ signal comes
from the non-resonant $B$ decays. The measured yields lead to
$\mathcal{B}(B\rightarrow
X(K\pi)_{non-res})\cdot\mathcal(X\rightarrow
J/\psi\pi\pi)=(8.1\pm2.0^{+1.1}_{-1.4})\times 10^{-6}$ and
$\mathcal{B}(B\rightarrow XK^*)\cdot\mathcal(X\rightarrow
J/\psi\pi\pi)<3.4\times 10^{-6}$ at 90\% C.L. The ratio
between two $B$ decays is
opposite to the one seen in a case of well known conventional
$c\overline{c}$ states where the resonant decay through $K^*$ is
dominant over the non-resonant $K^+\pi^-$ and 
adds another puzzle to the nature of $X(3872)$.

A second important update concerns decays of the $X(3872)$ into
$D^0\overline{D}^{*0}$ final state. The measured mass in
this final state is one of the driving points of a four-quark
hypothesis. On the other hand, measuring mass in this decay
mode is a non-trivial task due to the proximity of
kinematical
threshold. The Belle experiment provided update, using 657
million $B\overline{B}$ pairs \cite{Adachi:2008su} to
measure mass of $M=3872.6^{+0.5}_{-0.4}\pm 0.4$ \mevcc. 
Corresponding older measurement from \babar experiment yields
the mass $M=3875.1^{+0.7}_{-0.5}\pm 0.5$ \mevcc
\cite{Aubert:2007rva}. It is worth to note that the new Belle
mass measurements are compatible between
$\Jpsi\pi^+\pi^-$ and $D^0\overline{D}^{*0}$ decay modes.
While a tetraquark interpretation cannot be excluded based on
these measurements, the new mass measurement in
the $D^0\overline{D}^{*0}$ and small $\delta M$ in
the $X(3872)\rightarrow \Jpsi\pi^+\pi^-$ decay mode disfavor
particular model of Maiani et al \cite{Maiani:2007vr}.

\begin{figure}
\centering
\includegraphics[height=0.25\textwidth]{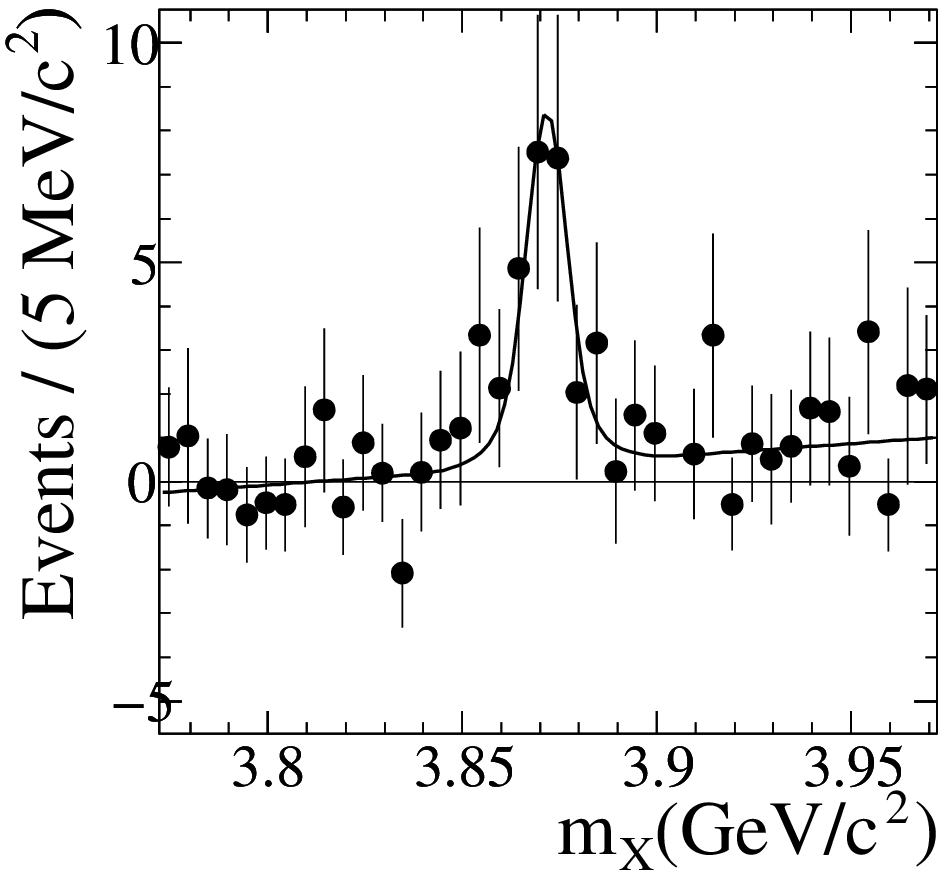}
\hspace{0.5cm}
\includegraphics[height=0.25\textwidth]{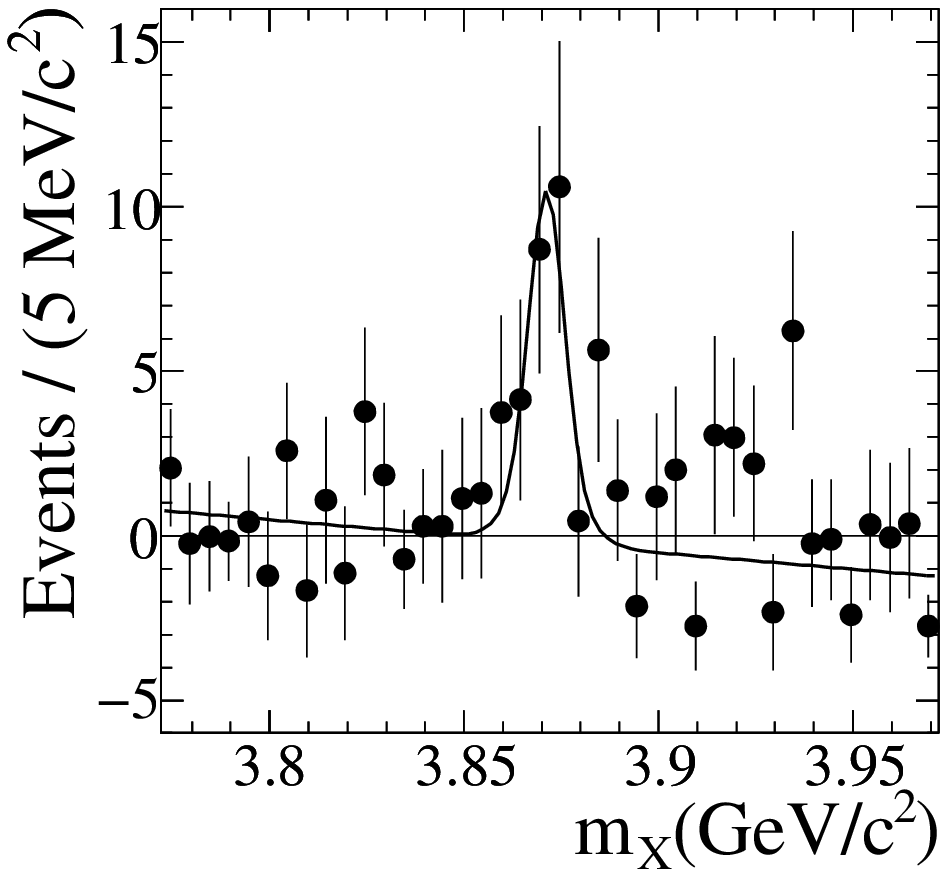}
\caption{The $\Jpsi\gamma$ (left) and $\psi(2S)\gamma$
(right) invariant mass distribution of $B^+$ signal events
from \babar experiment.}
\label{fig:babar_x3872_psig}
\end{figure}
A third measurement to mention is the search for radiative
decays of the $X(3872)$ to charmonium states. The \babar
experiment presented recent study of decays
$X(3872)\rightarrow \Jpsi\gamma$ and $X(3872)\rightarrow
\psi(2S)\gamma$ using their full dataset
\cite{Aubert:2008rn}. An observation of
radiative decays will fix charge parity on one hand and
a pattern of branching fractions to different $c\overline{c}$
states can distinguish different interpretations on the other
hand. While $D^0\overline{D}^{*0}$ molecule hypothesis can
accommodate decays to the $\Jpsi\gamma$, decays to
the $\psi(2S)\gamma$ are expected to be very small in such a case.
The obtained $\Jpsi\gamma$ and $\psi(2S)\gamma$ invariant mass
distributions are shown in Fig.~\ref{fig:babar_x3872_psig}.
Both decay modes show evidence for a signal with a
significance of about $3.5\sigma$.
The ratio of branching fractions is measured to
be $\mathcal{B}(X(3872)\rightarrow
\psi(2S)\gamma)/\mathcal{B}(X\rightarrow
\Jpsi\gamma)=3.4\pm1.4$, which indicates that even if
the $X(3872)$ is a $D^0\overline{D}^{*0}$ molecule, it has
a significant $c\overline{c}$ component.

\section{States around $3940$ MeV}

\begin{figure}
\centering
\includegraphics[width=0.45\textwidth]{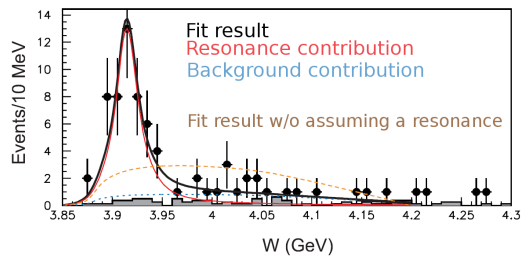}
\caption{The number of $\Jpsi\omega$ events as a function of
$\gamma\gamma$ center of mass energy from Belle experiment.}
\label{belle_X3915}
\end{figure}

A second area with recent result concerns states seen in
the $\Jpsi\omega$ final state with masses around $3940$ \mevcc.
History of this region goes back to 2005, when Belle
using $B$ decays observed a signal close to the $\Jpsi\omega$
threshold \cite{Abe:2004zs}. The measured resonance parameters are
$M=3943\pm 11\pm 13$ \mevcc and $\Gamma=87\pm 22\pm 26$
\mevcc. Later the signal was confirmed by \babar experiment,
again in B decays with parameters
$M=3914.3^{+3.8}_{-3.4}\pm2.0$ \mevcc and
$\Gamma=34^{+12}_{-8}\pm{ 5.0}$ \mevcc \cite{Aubert:2007vj}. Recently,
Belle experiment a performed study of the $\Jpsi\omega$ final
state in the $\gamma\gamma$ fusion. The preliminary invariant mass
distribution is shown in Fig.~\ref{belle_X3915}. A clear signal with
$55\pm14^{+2}_{-14}$  events and a significance of
$7.7\sigma$ is observed. The fit returns
$M=3914\pm 3\pm 2$ \mevcc and $\Gamma=23\pm 10^{+2}_{-8}$
\mevcc. 
It is currently unclear, whether three signals seen in the
$\Jpsi\omega$ final state are due to single resonance or
not.
Additionally it is also unclear whether we see
a new exotic state, or whether at least some of the peaks are actually
different decay mode of the conventional $c\overline{c}$
state $\chi_{c2}'$.

\section{The $1^{--}$ $Y$ family}

Next puzzling area consist of four states observed in the ISR
production and decaying to the $\Jpsi\pi\pi$ or
$\psi(2S)\pi\pi$ \cite{ISR1}. The four observed states are
named $Y$. The masses of the two states
seen in $\Jpsi\pi\pi$ are about 4008 and 4250 \mevcc and
those seen in the $\psi(2S)\pi\pi$ final state about 4360
and 4660 \mevcc. The widths of the states are about 100
\mevcc
with narrowest one being 48 \mevcc. The \babar experiment
presented a recent update of the measurement in
$e^+e^-\rightarrow \Jpsi\pi\pi \gamma_{ISR}$. The updated
values of the resonance parameters of $Y(4260)$ are
$M=4252 \pm6~^{+2}_{-3}$ \mevcc and $\Gamma=105 \pm
18^{+4}_{-6}$ \mevcc. It should be noted, that \babar
experiment observes only $Y(4260)$ and $Y(4360)$ states
without convincing signal for the other two. While for the
$Y(4008)$ even Belle signal is not very convincing, for
$Y(4660)$ the Belle signal is very clear. Also \babar
experiment sees a small enhancement around 4660 \mevcc in
$\psi(2S)\pi\pi$ final state, which can be interpreted
as resonance whose parameters are consistent with Belle
result, but the excess does not pass $3\sigma$ significance.
Also search in the open
charm meson pairs was performed. The Belle experiment
studied $DD$ final state \cite{Pakhlova:2008zza} while new \babar experiment
measures three different $D^{(*)}D^{(*)}$ final states
\cite{Aubert:2009xs}. In both cases, none of the $Y$ states can be
identified with a clear signal in the mass distributions of
$D^{(*)}D^{(*)}$ pairs.

The four states discussed here have $J^{PC}=1^{--}$ which is
fixed by the production mechanism in which they are observed. 
There seems to be no evidence for a decay to open charm
mesons as expected for a conventional $c\overline{c}$ states
above the open charm threshold. If all four states will be
confirmed with higher statistics experiment, then it will be
difficult to accommodate all of them into convention
charmonium spectrum as there are not enough unobserved
states available.

\section{Resonances in $\Jpsi\phi$}

All $XYZ$ states observed so far decayed to final
states which does not contain strange quarks. The situation
changed early this year, when CDF collaboration using
$B^+\rightarrow \Jpsi\phi K^+$ decays published evidence for
a near threshold resonance in the $\Jpsi\phi$ channel, which was
named $Y(4140)$ \cite{Aaltonen:2009tz}. Using events within $\pm3$
resolutions around the $B^+$ mass, which contains $75\pm10$
$B^+$ signal events on a low combinatorial background, they
extract $14\pm5$ signal events in the narrow $\Jpsi\phi$
resonance. Fit with relativistic s-wave Breit-Wigner
function for the signal, three body phase space for the background
returns $M=4143.0\pm2.9\pm1.2$ \mevcc and
$\Gamma=11.7^{+8.3}_{-5.0}\pm3.7$ \mevcc. The estimated branching
fraction is
$\mathcal{B}(B^+\rightarrow Y(4140)K^+,Y\rightarrow
J/\psi\phi)=9.0\pm3.4\pm2.9\times 10^{-6}$.

\begin{figure}
\begin{minipage}[c]{0.45\textwidth}
\includegraphics[width=0.92\textwidth]{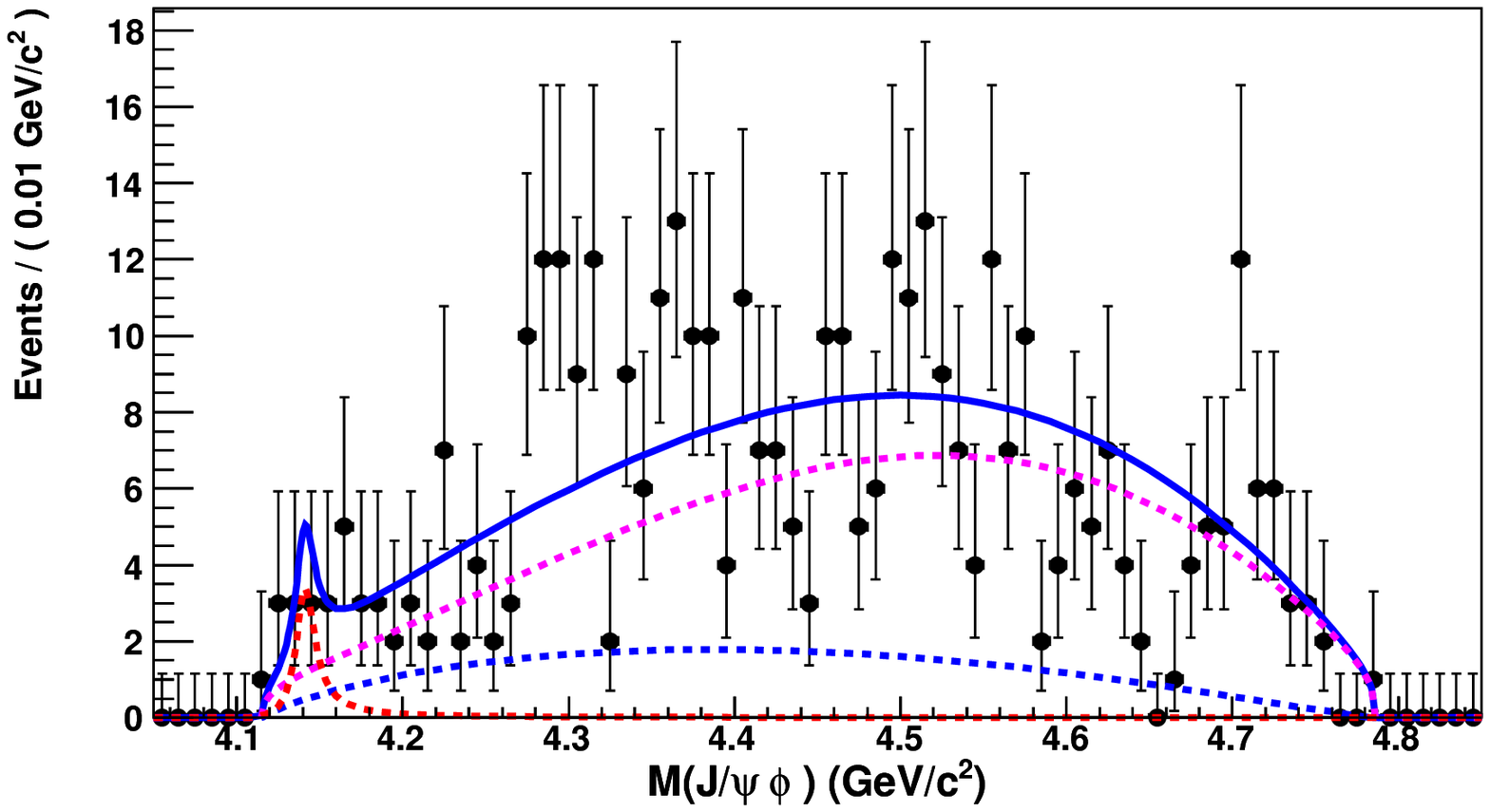}
\includegraphics[angle=90,width=0.85\textwidth]{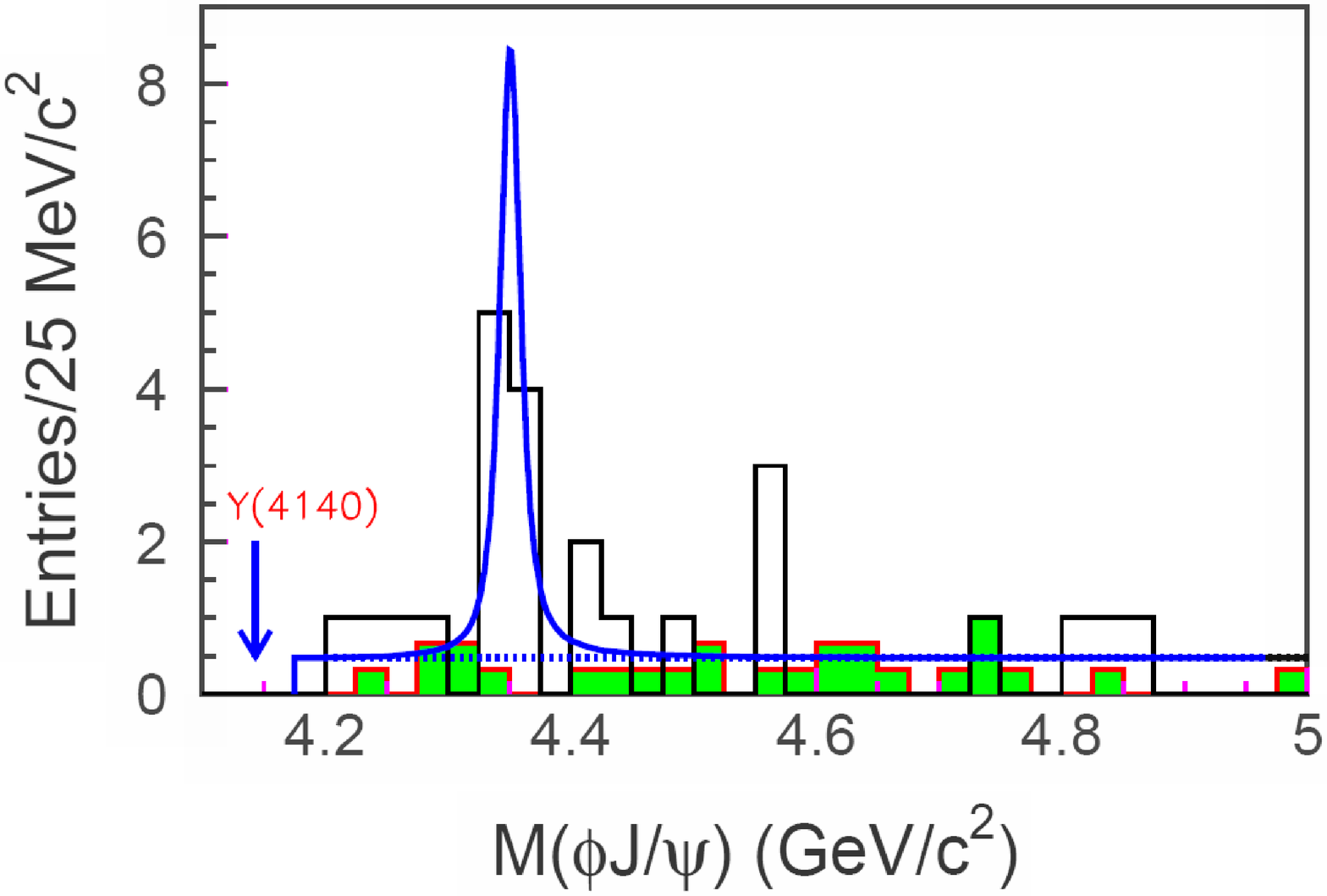}
\caption{The $\Jpsi\phi$ invariant mass distribution
obtained by Belle experiment in $B$ decays (top) and $\gamma\gamma$
fusion (bottom). The full line on top represents fit
projection, the magenta line $B^+$ background, the
dashed blue line non-$B^+$ background and the red
line signal. In bottom, the open histogram shows data, the
filled histogram $\Jpsi$ and $\phi$ sidebands
and the blue line fit projection.}
\label{belle_y4140}
\end{minipage}
\hfill
\begin{minipage}[c]{0.45\textwidth}
\includegraphics[width=0.95\textwidth]{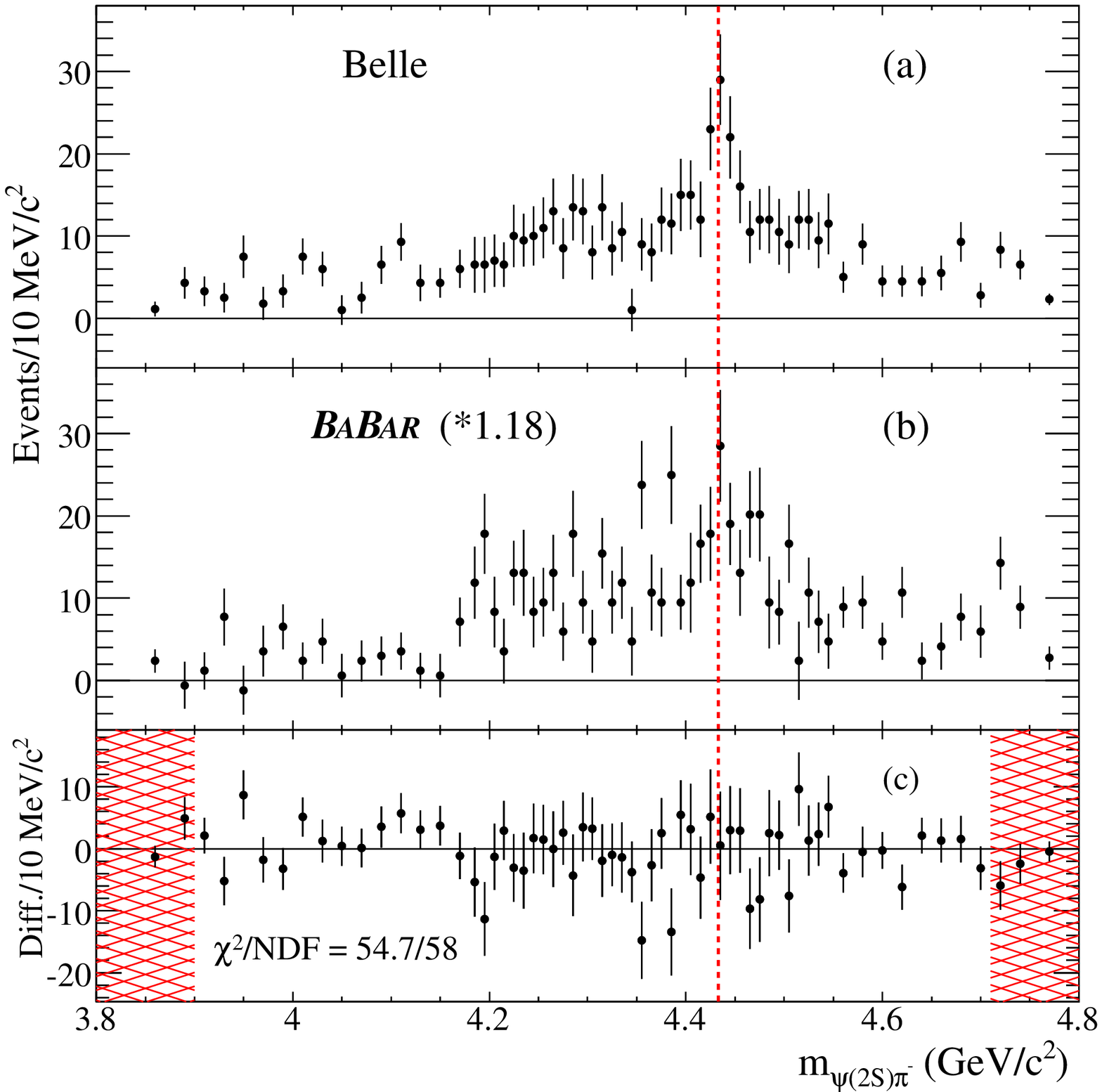}
\caption{The $\psi(2S)\pi^+$ invariant mass distributions of
$\overline{B}^0\rightarrow \psi(2S)\pi^+K^-$ signal from
Belle (top), \babar (middle) and difference of the two
(bottom). The distribution from \babar is scaled by factor
1.18 to take to account difference in the number of
$B\overline{B}$ pairs.}
\label{Z4430}
\end{minipage}
\end{figure}

The Belle experiment performs two different searches which
are sensitive to a resonance like one observed by CDF. 
First search is analogous to the CDF search, using
$B^+\rightarrow \Jpsi\phi K^+$ decays while second is search
in $\gamma\gamma\rightarrow \Jpsi\phi$ reaction. 
Both searches use almost all available data. 
In Fig.~\ref{belle_y4140} we show
$\Jpsi\phi$ invariant mass distribution of both searches.
The search in $B^+$ decays works with $325\pm21$ $B^+$ signal events on
low background. No significant signal is seen. In case of
the resonance parameters fixed to the CDF values, we obtain
$7.5^{+4.9}_{-4.4}$ signal events and set an upper limit on
the branching fraction $\mathcal{B}(B^+\rightarrow
Y(4140)K^+,Y\rightarrow J/\psi\phi) < 6\cdot 10^{-6}$ at
90\% C.L.
The search in $\gamma\gamma$ fusion observes no events
around 4140 \mevcc and sets an upper limit on $\gamma\gamma$
width times branching fraction
$\Gamma_{\gamma\gamma}(Y(4140))\mathcal{B}(Y(4140)\rightarrow\Jpsi\phi)
< 39$ eV/$c^2$. While there is no direct way of comparing
results in $B^+$ decays and $\gamma\gamma$ fusion, this
result indicates that if $Y(4140)$ is real, it is improbable
that it would be $D_s^{*+}D_s^{*-}$ molecule.

Additional interesting fact of the mass spectrum of $\gamma\gamma$ fusion
search is the cluster of events around 4350 \mevcc.
The fit of the data in the range from 4.2 to 5.0 \gevcc yields
$8.8^{+4.2}_{-3.2}$ signal events with a $3.9\sigma$
statistical significance. If interpreted as resonance, mass
is $M=4350.6^{+4.6}_{-5.1}\pm 0.7$ \mevcc and width
$\Gamma=13.3^{+17.9}_{-9.1}\pm4.1$ \mevcc.

\section{Charged $Z$ states}

The last topic to discuss are charged $Z$ states. Charged
states have an unique feature that by construction they cannot be accommodated
into the conventional $c\overline{c}$ spectrum. 
Two different final states show positive result up to now.
First one was $\psi(2S)\pi^+$ where in the
$\overline{B}^0\rightarrow \psi(2S)\pi^+ K^-$ decays Belle
observed a peak at about 4430 \mevcc \cite{Choi:2007wga}.
Second positive observation is
from the $B^0\rightarrow \chi_{c1}\pi^+K^-$ with two resonances
in $\chi_{c1}\pi^+$ at masses of about 4050 and 4250 \mevcc,
observed by Belle \cite{Mizuk:2008me}. 

News consist of analysis of $\overline{B}^0\rightarrow
\psi(2S)\pi^+ K^-$ final state by both B-factories
experiments. First, analysis of \babar data was presented
\cite{Aubert:2008nk}. Performing a detailed study of
the acceptance and possible reflections they concluded that no
significant signal exists in the data. The most
significant excess is at mass $4476\pm8$ \mevcc with
a $2.7\sigma$ significance. The mass of this excess is slightly
higher than $4433\pm4\pm1$ \mevcc measured
at Belle and also shows up mainly in the $K^*$ regions of the Dalitz
plot. With the same $K^*$ veto as done in original Belle
analysis \cite{Choi:2007wga} and resonance parameters fixed
to Belle values, small excess with $1.9\sigma$ significance
is fitted. On the Belle side, original dataset was
reanalyzed using original selection and  employing a full Dalitz plot ansatz
\cite{Mizuk:2009da}. The new analysis confirms previous
result with resonance parameters
$M=4433^{+15}_{-12}~^{+19}_{-13}$ \mevcc and
$\Gamma=107^{+86}_{-43}~^{+74}_{-56}$ \mevcc. Main change
compare to the original result is an increase in uncertainties,
which comes mainly from the uncertainty in Dalitz model.
It is worth to note, that while two experiments made
different conclusion, the data itself seems to be in
a reasonable agreement. In Fig.~\ref{Z4430} we show
$\psi(2S)\pi^+$ invariant distribution of $B^0$ signal for
both experiments and their difference. As one can see, there
is no large discrepancy and it is perhaps only lower
available statistics of the \babar experiment, which does
not allow to observe the $Z(4430)$.

\section{Conclusions}

Over the past year both B-factories made progress on studies of
$XYZ$ states. The $X(3872)$ received additional attention
with several new results and remains the best studied state.
Despite the experimental progress, there seems to be no
large progress on understanding of the $X(3872)$, except of
decreased chance that state seen in $\Jpsi\pi^+\pi^-$ is
different from the one seen in $D^0\overline{D}^{*0}$. 
Situation concerning charged $Z$ states basically didn't
change, they are still seen only by Belle experiment.
\babar performed search for $Z(4430)$, but while data seems
to be consistent with Belle, they do not reveal evidence for
the state.
Other studies performed in the past year revealed some
additional information and some signals, but none of them
was able to help in the understanding of $XYZ$ states. Altogether
there are more states seen than expected and unobserved
$c\overline{c}$ states. So if all of them remain, at least
some have to be of exotic origin. The progress on experimental side
probably needs to wait for future facilities which can
perform studies with tenfold statistics.

\end{document}